\documentclass[aps,pre,preprint,superscriptaddress,showpacs]{revtex4-1}

\usepackage{graphicx}
\usepackage{amssymb,amsmath}

\usepackage{color}

\begin{document}

\title{Colloids in light fields: particle dynamics in random and periodic energy landscapes}

\author{F.~Evers}
\email[]{Florian.Evers@hhu.de}
\affiliation{Condensed Matter Physics Laboratory, Heinrich Heine University, 40225 D\"usseldorf, Germany}

\author{R.D.L.~Hanes}
\affiliation{Condensed Matter Physics Laboratory, Heinrich Heine University, 40225 D\"usseldorf, Germany}

\author{C.~Zunke}
\affiliation{Condensed Matter Physics Laboratory, Heinrich Heine University, 40225 D\"usseldorf, Germany}

\author{R.F.~Capellmann}
\affiliation{Condensed Matter Physics Laboratory, Heinrich Heine University, 40225 D\"usseldorf, Germany}

\author{J.~Bewerunge}
\affiliation{Condensed Matter Physics Laboratory, Heinrich Heine University, 40225 D\"usseldorf, Germany}

\author{C.~Dalle-Ferrier}
\affiliation{Condensed Matter Physics Laboratory, Heinrich Heine University, 40225 D\"usseldorf, Germany}

\author{M.C.~Jenkins}
\affiliation{Condensed Matter Physics Laboratory, Heinrich Heine University, 40225 D\"usseldorf, Germany}

\author{I.~Ladadwa}
\affiliation{Institut f\"ur Physikalische Chemie, University M\"unster, 48149 M\"unster, Germany}
\affiliation{Fahad Bin Sultan University, 71454 Tabuk, Saudi-Arabia}

\author{A.~Heuer}
\affiliation{Institut f\"ur Physikalische Chemie, University M\"unster, 48149 M\"unster, Germany}

\author{R.~Casta\~neda-Priego}
\affiliation{Division of Sciences and Engineering, University of Guanajuato, 37150 Le\'on, Mexico}

\author{S.U.~Egelhaaf}
\email[]{Stefan.Egelhaaf@hhu.de}
\affiliation{Condensed Matter Physics Laboratory, Heinrich Heine University, 40225 D\"usseldorf, Germany}

\begin{abstract} 
The dynamics of colloidal particles in potential energy landscapes have mainly been investigated theoretically.
In contrast, here we discuss the experimental realization of potential energy landscapes with the help of light fields and the observation of the particle dynamics by video microscopy.
The experimentally observed dynamics in periodic and random potentials are compared to simulation and theoretical results in terms of, e.g. the mean-squared displacement, the time-dependent diffusion coefficient or the non-Gaussian parameter.
The dynamics are initially diffusive followed by intermediate subdiffusive behaviour which again becomes diffusive at long times.
How pronounced and extended the different regimes are, depends on the specific conditions, in particular the shape of the potential as well as its roughness or amplitude but also the particle concentration.
Here we focus on dilute systems, but the dynamics of interacting systems in external potentials, and thus the interplay between particle--particle and particle--potential interactions, is also mentioned briefly.
Furthermore, the observed dynamics of dilute systems resemble the dynamics of concentrated systems close to their glass transition, with which it is compared.
The effect of certain potential energy landscapes on the dynamics of individual particles appears similar to the effect of interparticle interactions in the absence of an external potential.
\end{abstract}

\maketitle



\section{Introduction}
\label{intro}

The motion of colloidal particles in potential energy landscapes is a central process in statistical physics which is relevant for a variety of scientific and applied fields such as hard and soft condensed matter, nanotechnology, geophysics and biology  \cite{Bouchaud1990,Sahimi1993,Sancho2010}.
Particle diffusion in periodic and random external fields is encountered in many situations \cite{Haus1987,Havlin1987,Wolynes1992,Dean2007}, such as atoms, molecules, clusters or particles moving on a surface with a spatially varying topology or interaction \cite{Sancho2004}, or moving through inhomogeneous bulk materials, e.g.~porous media or gels \cite{Dickson1996}, rocks \cite{Sahimi1993}, living cells or biological membranes \cite{Seisenberger2001,Weiss2004,Tolic2004,Banks2005,Hoefling2013,Barkai2012}.
It also includes the diffusion of charge carriers in a conductor with impurities \cite{Bystrom1950,Heuer2005}, particle diffusion on garnet films \cite{Tierno2010,Tierno2009,Tierno2012} or diffusion in optical lattices \cite{Carminati2001,Siler2010}, superdiffusion in active media \cite{Douglass2012}, and vortex dynamics in superconductors \cite{Harada1996}.
Moreover, some processes are modelled by diffusion in the configuration space of the system, e.g.~the glass transition \cite{Heuer2008,Debenedetti2001,Angell1995,Poon2002,vanMegen1998,Heuer2005a} and protein folding \cite{Best2011,Dobson1998,Bryngelson1995}. 

Thermal energy drives the Brownian motion of colloidal particles \cite{Haw2002,Frey2005}.
In free diffusion, their mean-squared displacement $\langle \mathrm{\Delta} x^2(t) \rangle$ increases linearly with time $t$; $\langle \mathrm{\Delta} x^2(t) \rangle \propto t^{\mu}$ with $\mu = 1$.
Particle--external potential (as well as particle--particle) interactions can modify the dynamics significantly leading to $\mu \neq 1$ \cite{Hoefling2013,Klafter2005,Sokolov2012,Schmiedeberg2007,Emary2012,Lichtner2012,Herrera2009,Euan2012}.
Often the dynamics slow down; on an intermediate time scale subdiffusion ($\mu < 1$) is observed, while at long times diffusion is reestablished with a reduced (long-time) diffusion coefficient $D_{\infty}$.
Different theoretical models have been developed to describe particle dynamics in external potentials, including the random barrier model \cite{Bernasconi1979}, the random trap model \cite{Schmiedeberg2007,Haus1982}, the continuous time random walk \cite{Scher1973}, diffusion in rough and regular potentials \cite{Dean2007,Zwanzig1988,Dieterich1977,Reimann2002}, the Lorentz gas model \cite{Hoefling2006}, and diffusion in quenched-annealed binary mixtures \cite{Krakoviack2005}.
Typically, theories focus on the asymptotic long-time limit, which is often difficult to reach in experiments.
In contrast, less is known about the behaviour at intermediate times, where the transitions between the different regimes occur.
Furthermore, theoretical calculations have mainly been exploited to extract information from experimental data, while only recently have theoretical predictions been compared systematically with experiments \cite{Tierno2010,Tierno2009,Tierno2012,Sciortino2005,Hanes2012,Evers2013,Dalle-Ferrier2011,Ladadwa2013,Hanes2012a,Skinner2013,Ma2013}.

Here we thus focus on recent experimental results on the dynamics of colloidal particles in potential energy landscapes and their comparison to simulation and theoretical predictions.
A prerequisite for systematic experiments is the controlled creation of external potential energy landscapes.
This, for example, is possible due to the interaction of colloidal particles with light \cite{Ashkin1980,Jenkins2008a,Dholakia2011,Dholakia2008a,Bowman2013,Neumann2004,Molloy2002,Ashkin1997}.
The effect of light on particles with a refractive index different (typically larger) from the one of the surrounding liquid is usually described by two forces: a scattering force or `radiation pressure', which pushes the particles along the light beam, and a gradient force, which pulls particles toward regions of high light intensity.
A classical application of this effect is optical tweezers which are used to trap and manipulate individual particles or ensembles of particles by a tightly focused laser beam or several laser beams, respectively \cite{Molloy2002,Ashkin1997,Grier2003,Hanes2009}.
Extended light fields rather than light beams can be used to create potential energy landscapes.
Arbitrary light fields can be generated using a spatial light modulator \cite{Dholakia2011,Hanes2009} or an acousto-optic deflector \cite{Bowman2013,Neumann2004,Juniper2012}, while crossed laser beams \cite{Jenkins2008a}, diffusors \cite{Bewerunge2013} and other optical devices can be used to create particular high-quality light fields (Sec.~\ref{sec:light}).

Light fields can affect the arrangement and dynamics of colloidal particles within individual phases but can also induce phase changes.
For example, upon increasing the amplitude of a periodic light field applied to a colloidal fluid, a disorder-order transition is induced in a two-dimensional charged colloidal system, known as light-induced freezing \cite{Herrera2009,Wei1998,Loudiyi1992,Bechinger2000}.
A further increase of the amplitude results in the melting of the crystal into a modulated liquid; this process is called light-induced melting.  
Extended light fields can also be applied to direct heterogeneous crystallization and hence the structure and unit cell dimensions of the formed bulk crystals or quasi-crystals \cite{vanBlaaderen2003,Brunner2002,Tata2012,Jaquay2013,Mikhael2008}.
Using light fields, the effect of periodic as well as random potentials on the particle dynamics has been experimentally investigated \cite{Hanes2012,Evers2013,Dalle-Ferrier2011} and compared to simulation and theoretical predictions \cite{Dean2007,Schmiedeberg2007,Emary2012,Lichtner2012,Zwanzig1988,Dalle-Ferrier2011,Ladadwa2013,Hanes2012a,Dalle-Ferrier2013,Festa1978}.
Most of the theoretical predictions only concern the asymptotic long-time behaviour.
Possible links between the long-time behaviour and the intermediate dynamics, as observed in the experiments, are discussed \cite{Evers2013,Hanes2012a,Hanes2014}.
Furthermore, the dynamics of individual particles in sinusoidal potentials are showing similarities with the dynamics in glasses \cite{Dalle-Ferrier2011,Vorselaars2007}.
Inspired by this idea, in this review the dynamics of individual particles in different external potentials are compared to the dynamics of concentrated hard spheres \cite{vanMegen1998,Zangi2004,Cui2001}.
Energy landscapes are not only considered in the context of glasses, but random energy landscapes with a Gaussian distribution of energy levels of width $\varepsilon \approx {\mathcal{O}}(k_{\text{B}}T)$, where $k_{\text{B}}T$ is the thermal energy, seem to be relevant for proteins, RNA and transmembrane helices \cite{Hyeon2003,Janovjak2007}.
Moreover, the diffusion (or `permeation') of rodlike viruses through smectic layers can be described by the diffusion in a sinusoidal potential of amplitude $\varepsilon \approx k_{\text{B}}T$ \cite{Lettinga2007,Grelet2008}.


\section{Colloids in light fields: creation of potential energy landscapes}
\label{sec:light}


The optical force on a colloidal particle has been investigated extensively, in particular in the context of optical tweezers \cite{Dholakia2008a,Bowman2013,Neumann2004,Molloy2002,Ashkin1997,Grier2003,Dholakia2008,Ashkin1992,Ashkin1986,Kerker1969,Harada1996a,Svoboda1994,Barnett2006}.
We consider a transparent colloidal particle with a refractive index $n_{\text{c}}$ suspended in a medium with a smaller refractive index $n_{\text{m}}$, that is $n_{\text{c}}>n_{\text{m}}$, and begin with the case of a particle much larger than the wavelength of light.
In this case, the simple picture of ray optics applies.
If light is incident on a particle, it will be scattered and reflected.
While light arrives from only one direction, the scattered and reflected light travels in different directions.
Hence the direction of the light and accordingly the momenta of the photons are changed.
Due to conservation of momentum, an equal but opposite momentum change will be imparted on the particle.
The rate of momentum change determines the force on the particle, which acts in the direction of light propagation and might, e.g.~due to the astigmatism of the objective, also have effects outside the main beam \cite{Zunke2013}.
This is the so-called scattering force or, considering the photon `bombardment', the radiation pressure.

When hitting the particle, the light beam will also be refracted, that is the particle acts as a (microscopic) lens.
This, again, changes the direction of the beam and hence the momentum of the photons.
The resulting force pushes the particle toward higher light intensities, mainly into the centre of the beam.
This is the gradient force, which acts in lateral direction and gradients typically also exist in axial direction, e.g.~toward a focus.
This decomposition of the optical force into two components, the scattering and gradient forces, is done traditionally although both originate from the same physics.


If the particle with radius $R$ is much smaller than the wavelength of light, $\lambda$, that is in the so-called Rayleigh regime, the particle's polarizability is considered.
The electric field of the light induces an oscillating dipole in the dielectric particle, which re-radiates light.
This leads to the scattering force \cite{Ashkin1986,Kerker1969,Harada1996a,Svoboda1994}
\begin{eqnarray}
F_{\text{scatt}} = \frac{ \sigma n_{\text{m}}}{c} I_0 \;\;\; \text{with} \;\;\;
\sigma = \frac{128 \pi^5 R^6 }{3 \lambda^4} \left( \frac{m^2-1}{m^2+2} \right)^2  \; ,
\label{eq:scat}
\end{eqnarray}
where $I_0$ is the incident light intensity, $\sigma$ the scattering cross section of a spherical particle, $c$ the speed of light and $m=n_{\text{c}}/n_{\text{m}}$.

The incident light intensity is typically inhomogeneous, $I_0(\vec{r})$, which leads to a further (component of the) force acting on the particle.
An induced dipole in an inhomogeneous electric field experiences a force in the direction of the field gradient, the gradient force \cite{Harada1996a,Svoboda1994}
\begin{eqnarray}
F_{\text{grad}} = \frac{2 \pi \alpha}{c n_{\text{m}}} {\vec{\nabla}} I_0(\vec{r}) \;\;\; \text{with} \;\;\;
\alpha = n_{\text{m}}^2 R^3 \left ( \frac{m^2-1}{m^2+2} \right )  \; ,
\label{eq:forces}
\end{eqnarray}
where $\alpha$ characterises the polarizability of a sphere.
The gradient force pushes particles with $n_{\text{c}}>n_{\text{m}}$ towards regions of higher intensity.


In the experiments described in the following, the particles are of comparable size to the wavelength of light.
However, this case is much more difficult to model \cite{Bowman2013,Neumann2004,Tlusty1998,Bonessi2007} and will thus not be described here.


In optical tweezers, tightly focused laser light is used to trap particles.
In contrast, exploiting the gradient force, here, extended spatially modulated light fields are applied to create potential energy landscapes \cite{Jenkins2008a}.
The modulations in the potential are relatively weak such that typically particles are not trapped for long times, but only remain for some time in certain areas.
Since the light field acts on the whole volume of the particle, its volume has to be convoluted with the light intensity to obtain the potential felt by the particle.
Depending on the size of the particle and the modulation of the light field, the centre of the particle might thus be attracted to bright or dark regions \cite{Jenkins2008a}.
Furthermore, it is difficult to impose potentials with features smaller than the particle size.


Extended space- and also time-dependent light fields can be created using various optical devices, e.g.~holographic instruments based on a spatial light modulator (SLM) \cite{Dholakia2011,Grier2003,Hanes2009} or an acoustic-optic deflector (AOD) \cite{Juniper2012}.
Spatial light modulators use arrays of liquid-crystal pixels.
Each pixel imposes a modulation of the phase, amplitude or polarization, which can be externally controlled.
This allows creation of almost any light field, within the limits of the finite size, pixelation and modulation resolution of the SLM.
The latter result in a noise component in the light field.
This can be exploited to create random potentials.
It can also be avoided by cycling different realizations of the same light field but with different phases, with a refresh rate beyond the structural relaxation rate of the sample \cite{Hanes2012,Evers2013,Vieten2013}.
Furthermore, the dynamic possibilities of a holographic instrument can be improved by combining it with galvanometer-driven mirrors \cite{Hanes2009}.

A conceptually simple but more specialized set-up is based on a crossed-beam geometry, which yields a standing wave pattern, i.e.~a sinusoidally-varying periodic light field, within an overlying Gaussian envelope due to the finite size of the beams \cite{Jenkins2008a,Loudiyi1992,Bechinger2000,Chowdhury1985,Koehler2000,Wiegand2004}.
Moreover, optical devices, such as diffusors, can be used to generate special beam shapes like top-hat geometries or randomly-varying light fields \cite{Bewerunge2013}.


While the gradient force is exploited to impose extended modulated potentials, whose amplitude is typically controlled by the laser power, the scattering force or radiation pressure will also affect the sample.
The radiation pressure determines the distance of the particle from the cover slip, which will thus depend on the laser power.
Due to hydrodynamic interactions, the distance to the cover slip affects the diffusion of the particle, which typically is reduced compared to free diffusion \cite{Pagac1996,Leach2009,Sharma2010}.
The experimental data presented in the following are corrected for this effect.


\section{Dynamics of individual colloids in periodic and random potentials}
\label{sec:dynamics}

\begin{figure}
\centering
\resizebox{1\columnwidth}{!}{\includegraphics{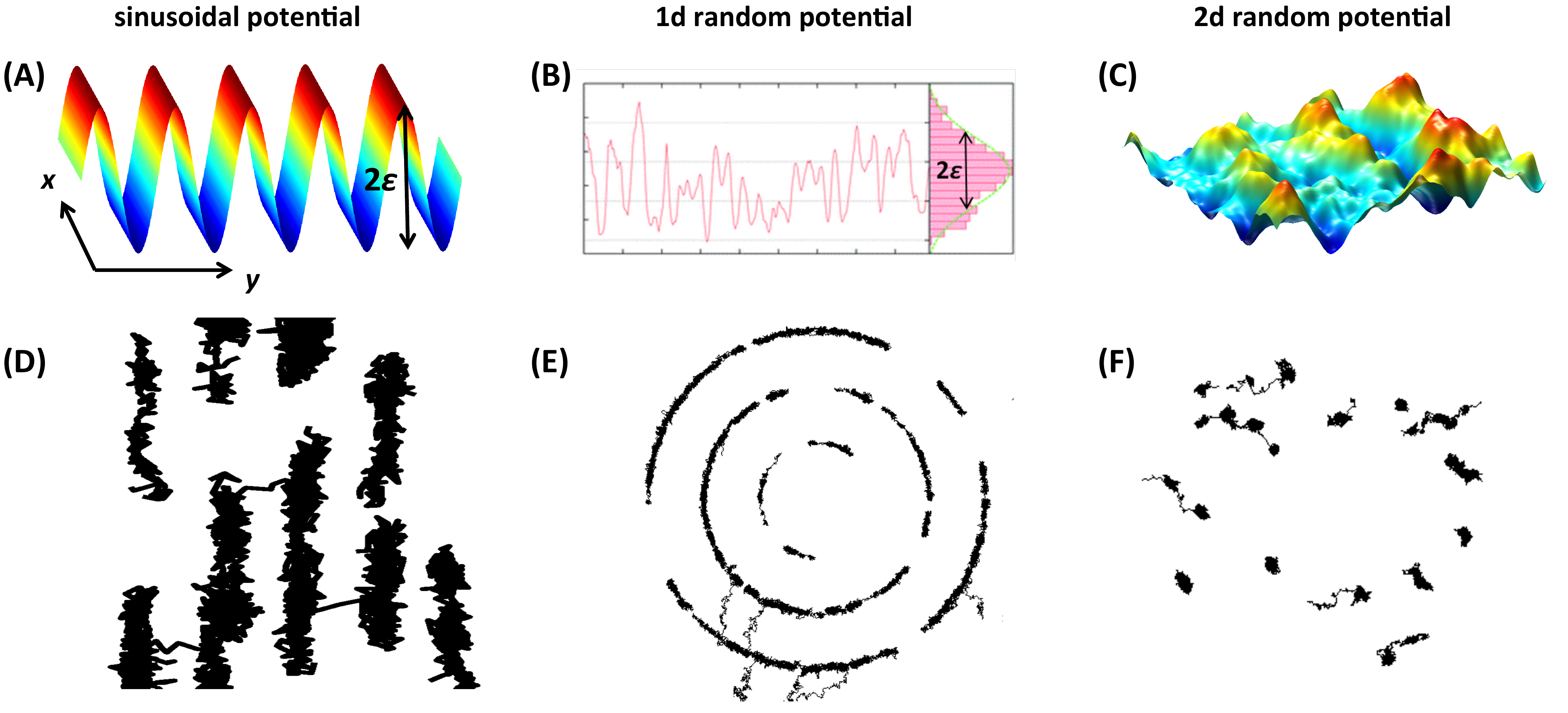}}
\caption{(top) Schematic representations of the potential energy landscapes as felt by the particles and as reconstructed from experimental data (left to right): sinusoidally-varying periodic potential \cite{Dalle-Ferrier2011}, one- and two-dimensional random potentials \cite{Hanes2012,Evers2013}. For the one-dimensional random potential, the histogram of values of the potential, $p(U)$, is shown and compared to a Gaussian distribution (green line).
(bottom) Representative particle trajectories in these potentials. The one-dimensional random potential was arranged in large circles to obtain `periodic boundary conditions' and to improve the statistics by simultaneously investigating several circles.} 
\label{fig:1}   
\end{figure}

Individual colloidal particles have been exposed to different potential energy landscapes (Fig.~\ref{fig:1}, top): sinusoidally-varying periodic potentials $U(y)=\varepsilon \sin{(2\pi y/\lambda)}$ with amplitude $\varepsilon$ and wavelength $\lambda$ (Fig.~\ref{fig:1}A), as well as one- and two-dimensional random potentials with a Gaussian distribution of potential values with (full) width $2 \varepsilon$ (Fig.~\ref{fig:1}B,C).
For the one-dimensional random potential, figure~\ref{fig:1}B shows the histogram of values of the potential, $p(U)$, which follows a Gaussian distribution $p(U) \propto {\exp}\{-(U-\langle U \rangle)^2/2 \varepsilon^2\}$.
In the experiments, the periodic potentials were generated using crossed laser beams \cite{Dalle-Ferrier2011,Jenkins2008a} and the random potentials using a spatial light modulator \cite{Hanes2012,Evers2013,Hanes2009} (Sec.~\ref{sec:light}).

The particle motions were monitored by video microscopy and the particle trajectories recovered by particle tracking algorithms \cite{Crocker1996,Jenkins2008}.
In the absence of a light field, i.e. without an external potential, colloidal particles undergo free diffusion, thus exploring large areas.
However, in the presence of external potentials, the particle dynamics are modified (Fig.~\ref{fig:1}, bottom).
The trajectories and hence the excursions of the particles were limited with the particles remaining for extended periods at positions that correspond to local minima of the potential.
In the periodic potential, anisotropic trajectories were observed (Fig.~\ref{fig:1}D).
Particle motion along the valleys ($x$ direction) was unaffected, while their motion across the maxima ($y$ direction) was hindered by barriers of height $2 \varepsilon$.

In the one-dimensional random potentials, the particles remained for different periods of time at different positions, reflecting the randomly-varying potential values along the circular path (Fig.~\ref{fig:1}E).
(The circular paths provided `periodic boundary conditions' and the use of several circles helped to improve statistics.)
Similarly, in the two-dimensional random potentials, the motion of the particles was limited due to the presence of local potential minima and saddle points (Fig.~\ref{fig:1}F).
Upon increasing the amplitude of the oscillations or the amplitude of the roughness, $\varepsilon$, the particles were more efficiently trapped and hence explored an even smaller region.


Based on the particle trajectories, different parameters were computed to characterize the particle dynamics quantitatively in the presence of external potentials.
The mean-squared displacement (MSD) is calculated according to
\begin{eqnarray}
\langle \mathrm{\Delta} x^2(t) \rangle
=  \left\langle \left[ x_i(t_0 + t) - x_i(t_0) \right]^2 \right\rangle_{t_0,i} \nonumber
- \left\langle \, \left[ x_i(t_0 + t) - x_i(t_0) \right] \, \right\rangle_{t_0,i}^2 \;\,  ,
\label{eq:msd2}
\end{eqnarray}
where the second term corrects for possible drifts.
For both, experiments and simulations, the average is taken over particles $i$ and waiting time $t_0$ to improve statistics.
The average over $t_0$ affects the results \cite{Ladadwa2013,Hanes2012a}, because initially the particles are randomly distributed while the distribution of occupied energy levels evolves toward a Boltzmann distribution.
To render the data independent of the specific experimental conditions, $\langle \mathrm{\Delta} x^2(t) \rangle$ was normalized by the square of the particle radius $R^2$, and the time $t$ by the Brownian time $t_{\text{B}}=R^2/(2 d D_0)$ with the short-time diffusion coefficient $D_0$ and the dimension $d$.

From the MSD, the normalized time-dependent diffusion coefficient $D(t)/D_0$ is calculated according to
\begin{eqnarray}
  \frac{D(t)}{D_0} = \frac{\partial \left ( \left\langle \mathrm{\Delta} x^2 (t) \right\rangle / R^2 \right )}{\partial (t/t_{\mathrm{B}})}   \;\; ,
  \label{eq:D}
\end{eqnarray} 
while the slope of the MSD in double-logarithmic representation
\begin{eqnarray}
  \mathrm{\mu}(t) = \frac{\partial \log{\left(\left \langle \Delta x^2(t) \right\rangle /R^2 \right)}}{\partial \log{(t/t_{\mathrm{B}})}   }
\end{eqnarray} 
corresponds to the exponent in the relation $\left\langle \mathrm{\Delta} x^2 (t) \right\rangle  \sim t^{\mathrm{\mu}(t)}$ and quantifies deviations from diffusive behaviour:
for free diffusion $\mathrm{\mu}=1$, while $\mathrm{\mu} < 1$ for subdiffusion and $\mathrm{\mu} > 1$ for superdiffusion.
In addition, the non-Gaussian parameter \cite{Vorselaars2007}
\begin{eqnarray}
  \mathrm{\alpha}_2(t) = \frac{ \left\langle \mathrm{\Delta} x^4 (t) \right\rangle}{(1+2/d)  \left\langle \mathrm{\Delta} x^2 (t) \right\rangle^2} -1
\end{eqnarray}
characterizes the deviation of the distribution of particle displacements from a Gaussian distribution and represents the first non-Gaussian correction \cite{Megen1998}.
In the two-dimensional case, the analogous equation based on $\langle \Delta r^2(t) \rangle$ and  $\langle \Delta r^4(t) \rangle$ was calculated and has the corresponding meaning.


\begin{figure}
\centering
\resizebox{1\columnwidth}{!}{\includegraphics{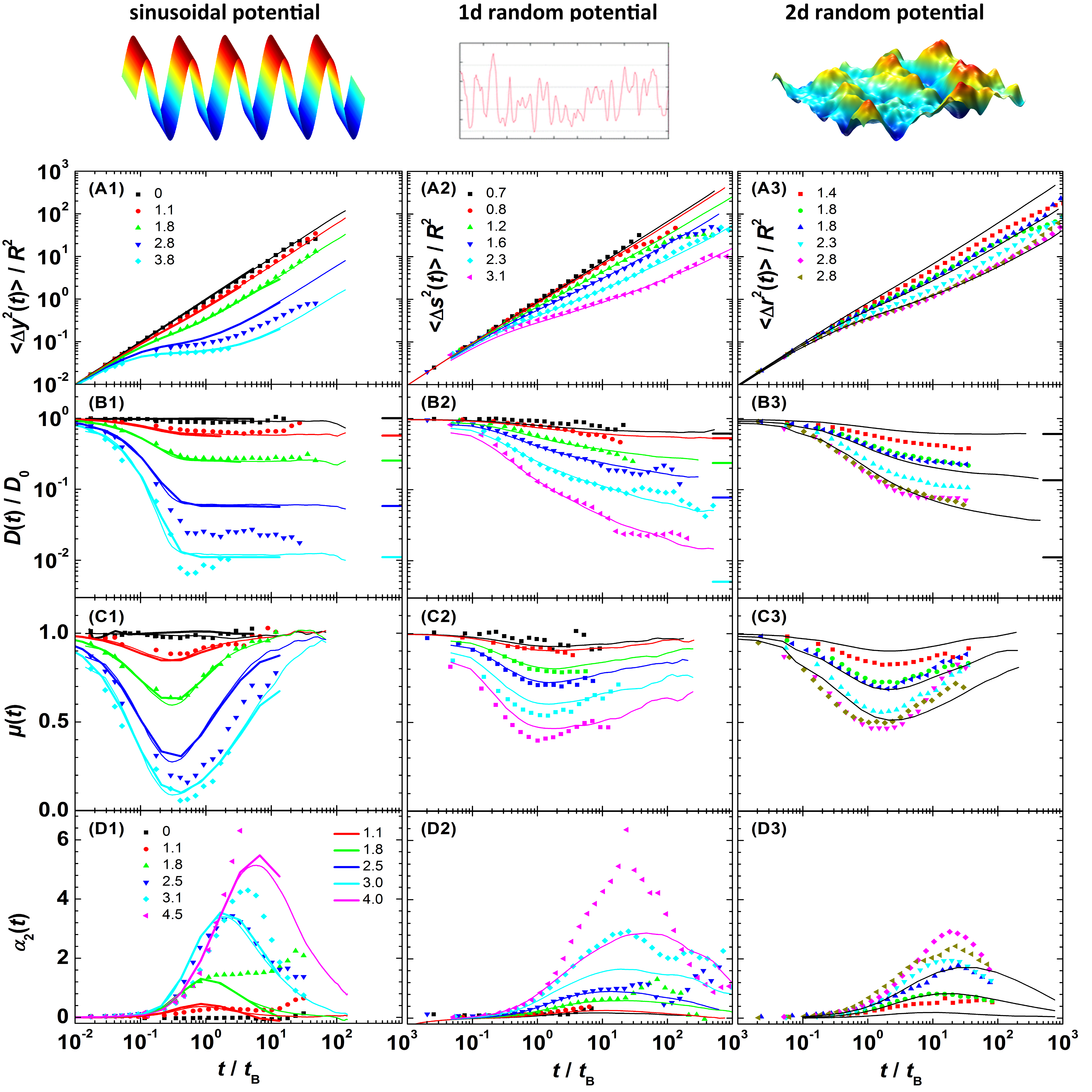}}
\caption{Particle dynamics in (left to right) sinusoidally-varying periodic \cite{Dalle-Ferrier2011}, one-dimensional random \cite{Hanes2012} and two-dimensional random potentials \cite{Evers2013} as characterized by (top to bottom) the normalized mean-squared displacements (where the normalization has been done according to the specific potential shape), the normalized time-dependent diffusion coefficient $D(t)/D_0$, the exponent $\mu(t)$ in the relation $\left\langle \mathrm{\Delta} x^2 (t) \right\rangle \propto t^{\mu(t)}$, and the non-Gaussian parameter $\alpha_2(t)$ for different potential amplitudes and degrees of roughness $\varepsilon$ (as indicated in the legends, in units of $k_{\text{B}}T$). Experimental data are represented by symbols, simulations by solid lines, theoretical predictions (for the periodic potential) by thick lines. Theoretical predictions for $D_\infty/D_0$ are indicated by horizontal lines.}
\label{fig:2}   
\end{figure}

The effect of potential shape and amplitude on the particle dynamics was investigated in experiments \cite{Hanes2012,Evers2013,Dalle-Ferrier2011}, simulations \cite{Ladadwa2013,Hanes2012a} and theory \cite{Dalle-Ferrier2011,Festa1978}, which all show consistent results (Fig.~\ref{fig:2}).
Without an external potential ($\varepsilon=0$), the MSD increases linearly with time and the diffusion coefficient $D(t)/D_0\approx 1$, exponent $\mu(t)\approx 1$ and non-Gaussian parameter $\alpha_2(t)\approx 0$ are all independent of time, as expected for free diffusion.
In contrast, in the presence of a periodic or random potential, the particle dynamics exhibit three distinct regimes which will be discussed in turn in the following.
(Note that in the case of the sinusoidal potential, we only discuss the motion across the barriers, i.e.~in $y$ direction.)


At short times, the particle dynamics are diffusive and follow the potential-free case.
This reflects small excursions within local minima and hence shows no significant dependence on the amplitude $\varepsilon$.

\begin{figure}
\centering
\resizebox{0.75\columnwidth}{!}{\includegraphics{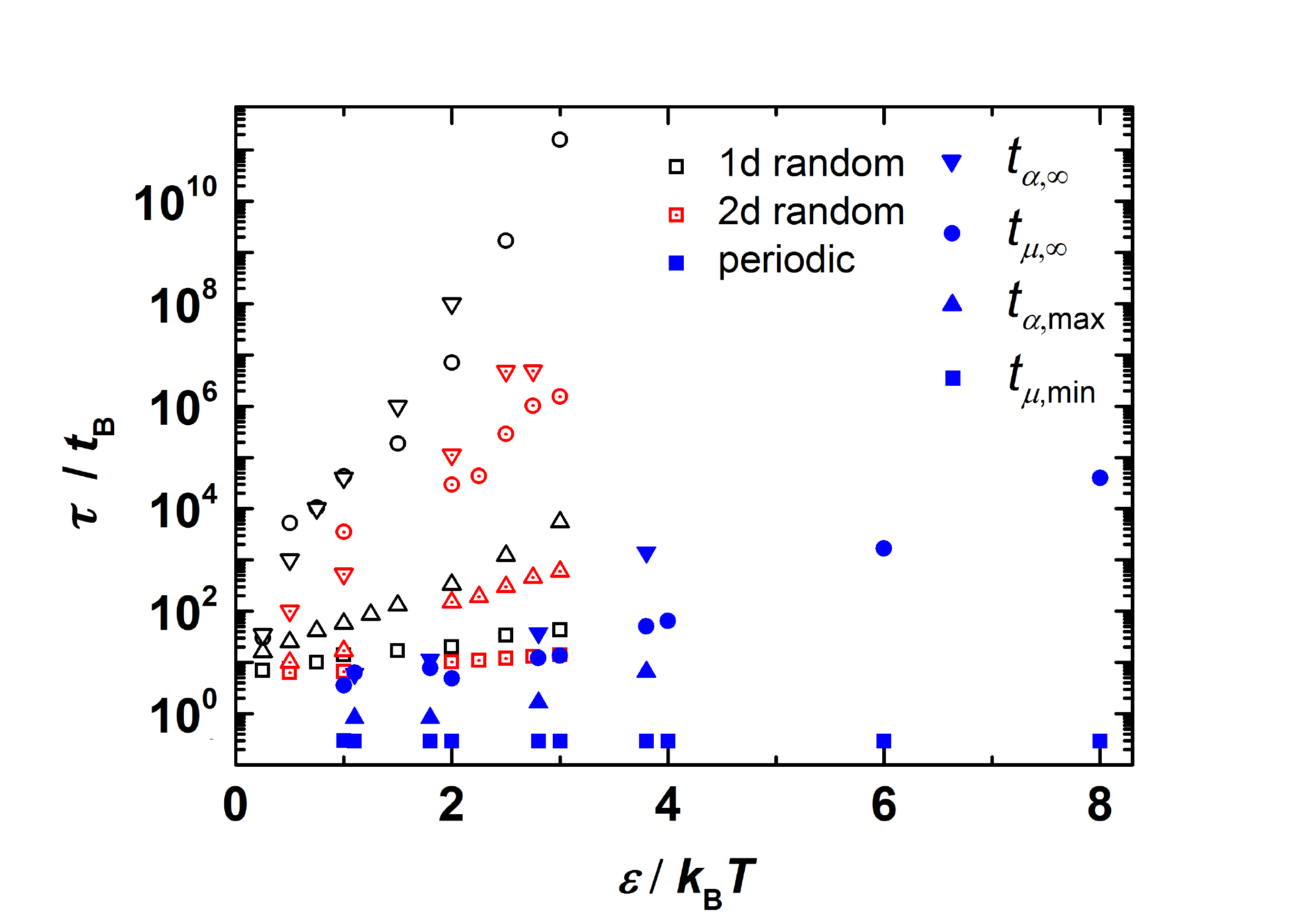}}
\caption{Characteristic time scales $\tau$ in sinusoidally-varying periodic, one-dimensional and two-dimensional random potentials (as indicated and explained in the text). Data for the periodic potential are extracted from theoretical results. For the random potentials, data are retrieved from simulation data not averaged over waiting times $t_0$ \cite{Ladadwa2013,Hanes2012a}.}
\label{fig:4}   
\end{figure}


At intermediate times, the MSDs exhibit inflection points or plateaux, which become increasingly pronounced as $\varepsilon$ increases.
This corresponds to the decrease of the diffusion coefficients $D(t)/D_0$ from 1 to significantly lower values, the decrease of the exponent $\mu(t)$ and the increase of the non-Gaussian parameter $\alpha_2(t)$.
The subdiffusive behaviour is caused by the particle being trapped in local minima for extended periods before it can escape to a neighbouring minimum.

In the case of the periodic potential, the barriers are all of equal height, $2\varepsilon$, and thus the residence time distribution is relatively narrow.
This is reflected in the reduced MSDs, the very pronounced and relatively quick decrease in $D(t)/D_0$ and $\mu(t)$ and increase in $\alpha_2(t)$.
Thus, the minima in $\mu(t)$ and maxima in $\alpha_2(t)$ occur at relatively short times, $t_{\mathrm{\mu,min}}$ and $t_{\mathrm{\alpha,max}}$, respectively (Fig.~\ref{fig:4}, blue solid symbols).
The minima in $\mu(t)$ occur earlier than the maxima in $\alpha_2(t)$.
This is due to the fact that the minimum in $\mu(t)$ reflects the largest deviation from diffusive behaviour, i.e.~when the probability to be (still) stuck in a minimum is largest and thus diffusion is most efficiently suppressed, whereas the maximum in $\alpha_2(t)$ indicates the largest deviation from the Gaussian distribution of displacements, i.e.~the dynamics are maximally heterogeneous with some minima having been left a long time ago, some only recently, with others yet to be left.
The maximum in $\alpha_2(t)$ thus only appears once jumps have occurred, which happens after the minimum in $\mu(t)$, and hence $t_{\mathrm{\mu,min}} < t_{\mathrm{\alpha,max}}$.
This also implies a weak $\varepsilon$ dependence of $t_{\mathrm{\mu,min}}$ and a significant $\varepsilon$ dependence of $t_{\mathrm{\alpha,max}}$ since $\varepsilon$ determines the height of the barrier which has to be crossed.
Similarly, the intermediate regime ends once the particle escapes the minima and performs a random walk between different minima with the diffusion coefficient $D(t)/D_0$, $\mu(t)$ and $\alpha_2(t)$ reaching the plateaux values, unity and zero, respectively.
Again, since all barrier heights are identical, this occurs within a short period of time.
Nevertheless, the time required to reach the end of the intermediate regime and hence the long-time diffusive limit, quantified either by $\mu\to 1$, i.e.~$t_{\mathrm{\mu,\infty}}$, or by $\alpha_2\to 0$, i.e.~ $t_{\mathrm{\alpha,\infty}}$, strongly depends on $\varepsilon$.

In the other case, i.e.~in the presence of a random potential, there exists a wide range of barrier heights and thus residence times.
This is reflected in the less pronounced pla\-teaux or rather inflection points in the MSDs, a very slow decrease of $D(t)/D_0$ with very slow approaches to the long-time plateaux as well as a slow decrease and increase of $\mu(t)$ and $\alpha_2(t)$, respectively, and in particular an extremely slow return of $\mu(t)$ and $\alpha_2(t)$ to 1 and 0, respectively.
Therefore, the intermediate subdiffusive regime, as indicated by the range from $t_{\mathrm{\mu,min}}$ and $t_{\mathrm{\alpha,max}}$ to $t_{\mathrm{\mu,\infty}}$ and $t_{\mathrm{\alpha,\infty}}$, occurs relatively late and in particular extends over a broad range of times with a strong $\varepsilon$ dependence (Fig.~\ref{fig:4}), where the particular $\varepsilon$ dependence of $t_{\mathrm{\mu,\infty}}$ and $t_{\mathrm{\alpha,\infty}}$ is still under debate \cite{Evers2013,Hanes2014}.
For the one-dimensional random potential, sub\-diffusion is more pronounced than for the two-dimensional random potential, since in two dimensions maxima can be avoided and only saddle points need to be crossed.
For the same reason, in one dimension, the $\varepsilon$ dependence appears stronger and the intermediate regime extends to longer times.
Thus, in the one-dimensional random potential the intermediate subdiffusive regime covers a longer time period than in the two-dimensional case, which in turn is longer and shows a stronger $\varepsilon$ dependence than in the periodic potential.
Moreover, increasing amplitude $\varepsilon$ has similar effects for all potential shapes:
First, the subdiffusive behaviour becomes more pronounced.
Second, the intermediate regime extends to longer times, indicated by the slow returns of $\mu(t)$ and $\alpha_2(t)$ to 1 and 0, respectively.
However, the beginning of the intermediate regime, characterized by the maxima in $\mu(t)$ and minima in $\alpha_2(t)$ and the corresponding times $t_{\mathrm{\mu,min}}$ and $t_{\mathrm{\alpha,max}}$, remains at about the same time with a weak $\varepsilon$ dependence since no or only a few barrier crossings are involved.
\footnote{Note that the amplitude $\epsilon$ characterises the amplitude of the oscillations in the case of the periodic potential, while it represents the amplitude of the roughness, namely the width of the Gaussian distribution of values of the potential, in the case of the random potentials.}
Extrapolations of the characteristic times $\tau$ to vanishing potential amplitudes results in different values $\tau(\varepsilon{\to}0)$ for the different potential shapes.
Although unexpected, this might be related to the definitions of the amplitude $\varepsilon$ for the periodic and random potentials, respectively, and to the fact that without an external potential, i.e.~$\varepsilon=0$, $\mu(t)=1$ and $\alpha_2(t)=0$ and thus no minimum in $\mu(t)$ and no maximum in $\alpha_2(t)$ exist and hence $\tau(\varepsilon{=}0)$ is not defined.

\begin{figure}
\centering
\resizebox{0.75\columnwidth}{!}{\includegraphics{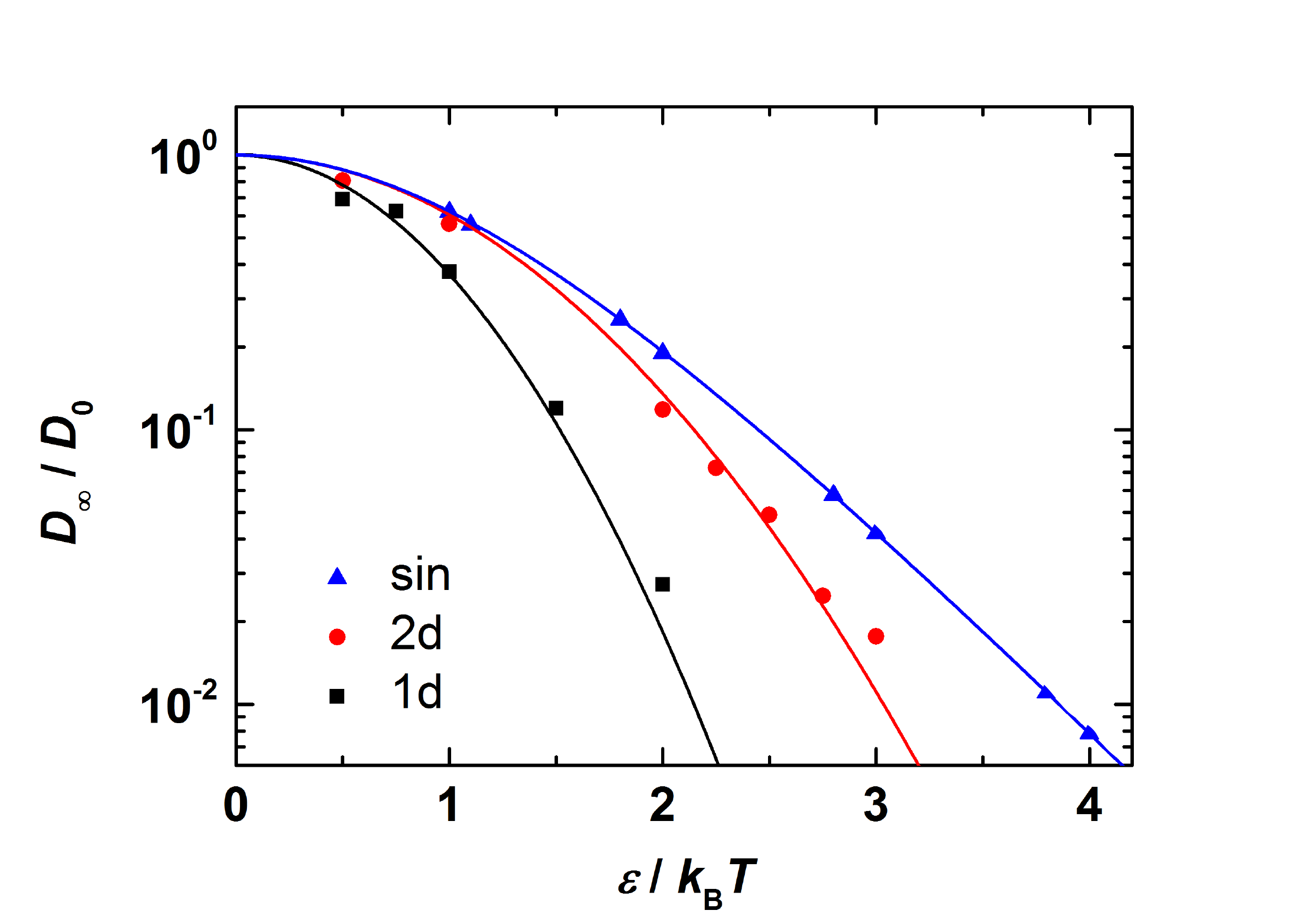}}
\caption{Normalized long-time diffusion coefficient $D_\infty(\varepsilon)/D_0$ in one- and two-dimensional random potentials and sinusoidally-varying periodic potentials (left to right). Solid lines indicate theoretical predictions \cite{Dean2007,Zwanzig1988,Festa1978}, symbols simulation results that have not been averaged over waiting times $t_0$ \cite{Dalle-Ferrier2011,Ladadwa2013,Hanes2012a}.
}
\label{fig:3}   
\end{figure}

At very long times, again diffusive behaviour is observed with constant, but much smaller $D_\infty/D_0$ and $\mu(t)$ returning to 1 and $\alpha_2(t)$ to 0.
On long time scales, hopping between minima becomes possible and, once more, the dominant process is a random walk, now between minima.
The return to diffusion is fast in the case of the periodic potential, since very deep minima are absent, but slow in the two- and especially the one-dimensional random potential.
With increasing amplitude $\varepsilon$, one notices increasingly long times to reach the asymptotic long-time limit (Fig.~\ref{fig:4}) and a decrease of the long-time diffusion coefficient $D_\infty(\varepsilon)$ (Fig.~\ref{fig:3}), which
has been calculated for different potential shapes.
For a periodic sinusoidal potential \cite{Dalle-Ferrier2011,Festa1978}
\begin{eqnarray}
   \frac{D_{\infty}(\varepsilon)}{D_0} = J_0^{-2}\left (\frac{\mathrm{\varepsilon}}{{k_{\text{B}}T}} \right) 
   \approx 2 \pi \left (\frac{\mathrm{\varepsilon}}{k_{\text{B}}T} \right) \mathrm{e}^{\left (\frac{-2 \mathrm{\varepsilon}}{k_{\text{B}}T} \right)} \; ,
   \label{eq:sin}
\end{eqnarray}
where $J_0$ is the Bessel function of the first kind of order $0$ and the approximation holds for $\varepsilon \gg k_{\text{B}}T/2$ \cite{Dalle-Ferrier2011}.
In the case of one- and two-dimensional random potentials one finds \cite{Dean2007,Zwanzig1988,Dean1997,Dean2004,Touya2007}
\begin{eqnarray}
   \frac{D_{\infty}(\varepsilon)}{D_0} = \mathrm{e}^{- \frac{1}{d} \left (\frac{\mathrm{\varepsilon}}{k_{\text{B}}T} \right)^2} \; .
   \label{eq:DD}
\end{eqnarray}
In the case of a two-dimensional random potential, $D_\infty(\varepsilon)$ is larger because maxima can be avoided and only saddle points have to be crossed.
Nevertheless, the exponential dependence on $-(\varepsilon/ k_{\text{B}}T)^2$ remains, which is the ratio of the equilibrium energy of a Gaussian distribution, $-\varepsilon^2/k_{\text{B}}T$, and the thermal energy $k_{\text{B}}T$.
The first term describes the equilibrium energy and dominates the dependence of the activation barriers on temperature, because the typical barrier energies to be overcome when moving between different regions are essentially independent of the thermal energy, as suggested by the percolation picture \cite{Dyre1995}.
The theoretical predictions and simulation as well as experimental data agree (Fig.~\ref{fig:3}), except at large $\varepsilon$ where deviations are noticeable.
Figure~\ref{fig:3} shows the theoretical predictions and simulation results, the latter agreeing with the experimental data (Fig.~\ref{fig:2}).
The slightly higher data are due to the fact that even for the longest investigated times the asymptotic long-time limit is not quite reached for the largest $\varepsilon$ (Fig.~\ref{fig:2}).
Moreover, the data suggest that the time to reach the long-time limit, characterised by $t_{\mathrm{\mu,\infty}}(\varepsilon)$ or $t_{\mathrm{\alpha,\infty}}(\varepsilon)$ (Fig.~Ê\ref{fig:4}), is not related to $D_\infty(\varepsilon)^{-1}$ (Fig.~\ref{fig:3}).

\begin{figure}
\centering
\resizebox{1\columnwidth}{!}{\includegraphics{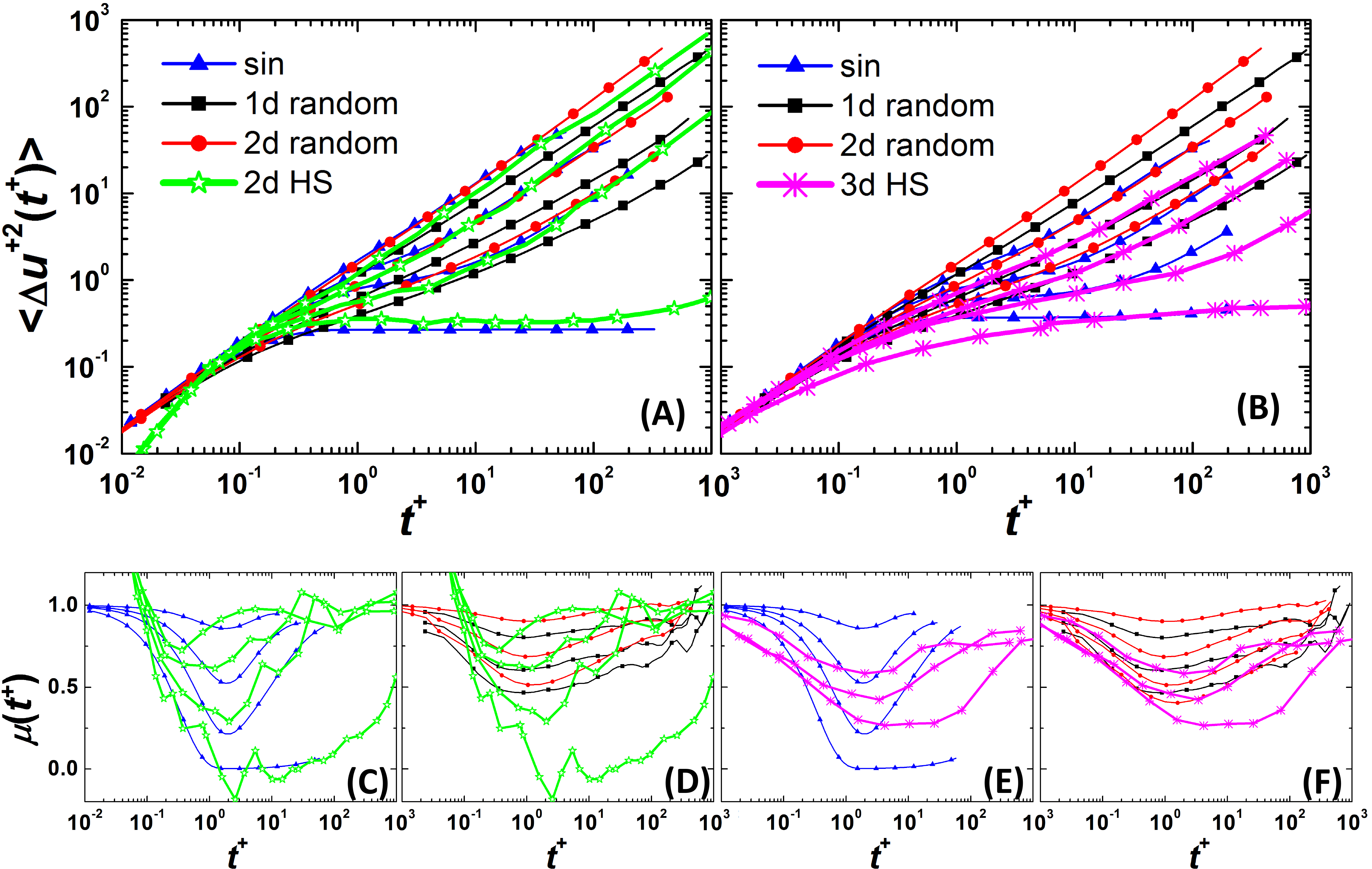}}

\caption{Effect of particle--particle and particle--potential interactions on the particle dynamics. The dynamics of individual particles in sinusoidally-varying periodic and one- and two-dimensional random potentials (thin lines with symbols as indicated) \cite{Hanes2012,Evers2013,Dalle-Ferrier2011} are compared to (A,C,D) (quasi) two-dimensional concentrated hard discs \cite{Zangi2004} and (B,E,F) three-dimensional concentrated hard spheres \cite{vanMegen1998}, the latter two in the absence of an external potential (thick lines as indicated). (A,B) Dimensionless mean squared displacements and (C--F) exponent $\mu(t)$ as function of dimensionless time. To allow for a comparison, both the mean squared displacement and the time have been normalized by typical length scales of the corresponding systems, indicated by the parameters $\langle \Delta u^{_{+}\,2}(t^+) \rangle$ and $t^+$. Shown are the theoretical predictions for individual particles in periodic potentials with amplitude (A,C) $\varepsilon / k_{\text{B}} T = 1, 2, 3, 8$ and (B,E) $2, 3, 4, 6$ \cite{Dalle-Ferrier2011}, simulation results for individual particles in one-dimensional random potentials with amplitude $\varepsilon / k_{\text{B}} T = 1.2, 2.3, 3.1$ \cite{Hanes2012a} and in two-dimensional random potentials with amplitude $\varepsilon / k_{\text{B}} T = 1.0, 2.0, 3.0$ \cite{Evers2013,Ladadwa2013}, simulation results for concentrated hard discs with surface fractions $\sigma=0.68, 0.69, 0.70, 0.715$ in the absence of an external potential \cite{Zangi2004}, experimental data for concentrated hard spheres with volume fractions $\Phi=0.466, 0.519, 0.558, 0.583$ in the absence of an external potential \cite{vanMegen1998} (all top to bottom).}
\label{fig:5}   
\end{figure}

The particle dynamics in periodic and random potentials as discussed above, resemble the dynamics of concentrated systems, whose subdiffusive behaviour has been associated with caging by neighbouring particles \cite{Pusey1986,Schweizer2007,Schweizer2003}.
Thus particle--potential and particle--particle interactions seem to have similar effects on the particle dynamics.
Their effects lead to characteristic signatures especially in the intermediate regime, which was described above.
We hence can compare the dynamics of individual particles in external potentials and concentrated interacting particles without external potential (Fig.~\ref{fig:5}), namely experimental data from a three-dimensional bulk system containing hard spheres of different volume fractions \cite{vanMegen1998} and experimental as well as simulation data from (quasi) two-dimensional systems of hard discs with different surface fractions \cite{Zangi2004,Cui2001}.
The dynamics of the concentrated two-dimensional system and the individual particles in the periodic potential are strikingly similar (Fig.~\ref{fig:5}A,C), while the dynamics in the random potentials appear different (Fig.~\ref{fig:5}A,D).
In contrast, the dynamics of the concentrated three-dimensional system seem different from the individual particles in the periodic potential, for example the intermediate MSD is broader (Fig.~\ref{fig:5}B,E), while it resembles the dynamics in the random potentials (Fig.~\ref{fig:5}B,F).

\section{Dynamics of interacting colloids in periodic and random potentials}
\label{sec:outlook}

\begin{figure}
\centering
\resizebox{0.75\columnwidth}{!}{\includegraphics{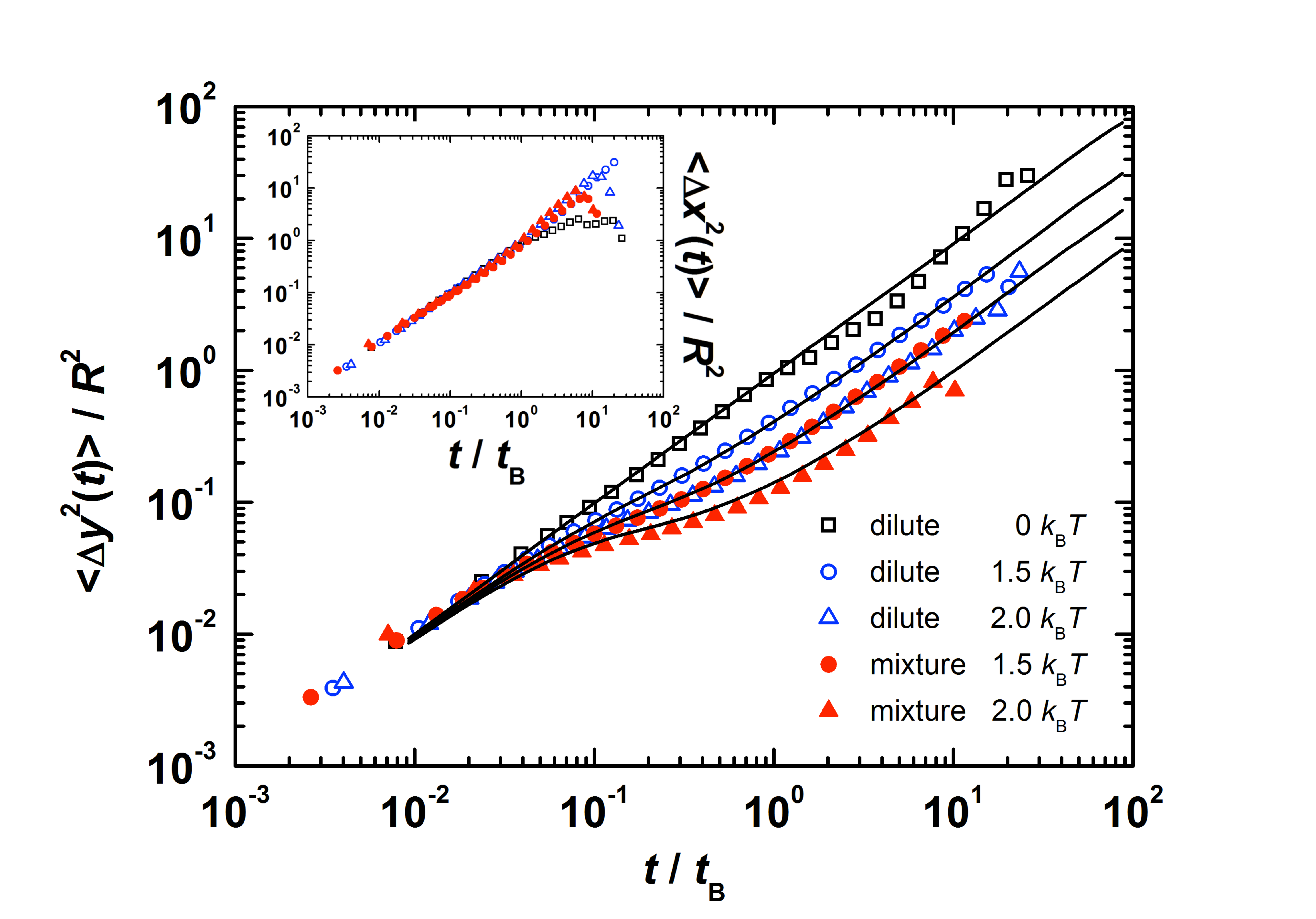}}
\caption{Particle dynamics, namely MSD across the barriers, i.e. in $y$ direction (main figure), and along the valleys, i.e. in $x$ direction (inset), of an individual dilute large particle ($R_1=2.5~\mu$m, open symbols) and concentrated large particles in a binary mixture ($R_1=2.5~\mu$m, $R_2=1~\mu$m, total surface fraction $\sigma \approx 0.57$ with an about equal number of large and small spheres, filled symbols), both in sinusoidally-varying periodic potentials with wavelength $\lambda=5.2~\mu$m and different amplitudes (as indicated). Lines represent Brownian Dynamics simulations of individual particles in a periodic potential with $\varepsilon/k_{\text{B}}T = 0.0, 1.5, 2.0, 2.5$.}
\label{fig:6}   
\end{figure}

So far the dynamics of individual colloidal particles in periodic and random potentials were considered.
It shows striking similarities with the dynamics of concentrated suspensions without external potentials \cite{vanMegen1998,Zangi2004,Cui2001}.
The combined effect of particle--potential and particle--particle interactions is hence briefly discussed.
An increase of the particle concentration in a one-dimensional channel leads to single file diffusion with $\langle \Delta x^2(t) \rangle \sim t^{0.5}$ \cite{Lutz2004}, which becomes more complex if a periodic potential is present along the channel \cite{Euan2012}.
Also in two-dimensional potentials an interplay between the particle--potential and particle--particle interactions was observed \cite{Herrera2009}, which, under the investigated conditions, may be linked to changes in the particle arrangement, caused by laser-induced freezing and melting \cite{Wei1998,Loudiyi1992,Bechinger2000}.
More complex potential-induced disorder-order and disorder-disorder transitions have been theoretically investigated in mixtures, namely colloid-polymer mixtures and binary hard discs \cite{Goetze2003,Franzrahe2007,Franzrahe2009}.
The dynamics of binary colloidal mixtures with large size disparity have been investigated without the presence of an external potential \cite{Imhof1995diff,Voigtmann2009}.
Here, we focus on the dynamics of concentrated binary hard sphere mixtures in a periodic potential, with the mixture in the modulated liquid state.
The MSDs of individual particles (similar to those in Sec.~\ref{sec:dynamics}) and of interacting particles in the presence of smaller particles in a periodic potential are determined  (Fig.~\ref{fig:6}).
No signature of single-file diffusion could be observed in the MSDs along the valleys, i.e. in $x$ direction (Fig.~\ref{fig:6}, inset).
Across the barriers, i.e.~in $y$ direction, the MSDs of the interacting large particles in the binary mixture (in a periodic potential with amplitude $\varepsilon$) resemble the MSDs of individual large particles (in a periodic potential with a larger amplitude $\varepsilon'$).
For the present conditions, in particular surface fraction $\sigma \approx 0.57$, we found $\varepsilon' \approx \varepsilon + 0.5~k_{\text{B}}T$.
Moreover, the MSDs of the individual and interacting particles in a periodic potential agree with Brownian Dynamics simulations of an individual particle in a periodic potential (Fig.~\ref{fig:6}, lines).
Similar observations have been made for interacting quasi-monodisperse particles in periodic and random potentials \cite{Dalle-Ferrier2013,Zunke2014}.

\section{Conclusion}
\label{conclucsion}

Optical devices, such as spatial light modulators and acousto-optic deflectors, can be exploited to create a large variety of modulated light fields.
Due to the polarizability of colloidal particles, this translates into potential energy landscapes of almost any shape.
The large flexibility, together with the possibility to observe and track colloidal particles by video microscopy, provides an ideal experimental tool to systematically and quantitatively investigate fundamental questions in statistical physics.
Here we focused on individual Brownian particles, but also briefly mentioned interacting particles, in periodic and random potentials.
The experimental findings were compared to simulation results and theoretical predictions.
While the latter mainly concerns the long-time asymptotic limit, the experiments and simulations also provide detailed quantitative information on the intermediate dynamics, which exhibit subdiffusive behaviour.
This was compared to the distinct intermediate dynamics of concentrated colloidal suspensions, thus comparing particle--potential and particle--particle interactions.
The interplay between these interactions was also illustrated using concentrated binary mixtures in external potentials.
The dynamics of concentrated interacting particles in potential energy landscapes deserve further work, which will also be extended to time-dependent potentials.

\section*{Acknowledgement}
We gratefully acknowledge support by the Deutsche Forschungsgemeinschaft (DFG) through the SFB-Transregio TR6 ``The Physics of Colloidal Dispersions in External Fields'' (project C7), the Research Unit FOR~1394 ``Nonlinear Response to Probe Vitrification'' and the International Helmholtz Research School `BioSoft'. C.~D.-F.~thanks the Humboldt foundation for the award of a fellowship. We thank H.E.~Hermes, J.~Horbach, M.~Laurati, H.~L\"owen, K.J.~Mutch, P.~Nielaba, K.~Sandomirski, M.~Schmiedeberg, D.~Wagner, A.~Yethiraj for very helpful discussions.

\bibliography{bib-v11}
 
\end{document}